\documentstyle[preprint,epsfig,aps]{revtex}
\draft
\begin{document}
\title{Brane cosmology with a bulk scalar field}
\author{David Langlois, Mar\'{\i}a Rodr\'{\i}guez-Mart\'{\i}nez}
\address{Institut d'Astrophysique de Paris, \\
(Centre National de la Recherche Scientifique)\\
98bis Boulevard Arago, 75014 Paris, France}
\date{\today}
\maketitle

\def\beq{\begin{equation}}
\def\eeq{\end{equation}}
\def\K52{{\kappa^2}}
\def\C{{\cal C}}
\def\A{{\cal A}}
\def\lamb{{\rho_{\Lambda}}}
\def\a{{\tilde \alpha}}
\def\F{{\tilde\phi}}
\def\U{{\tilde U}}
\def\V{{\tilde V}}
\def\d{{\delta}}
\def\g{{ \gamma}}

\begin{abstract}
We consider ``cosmologically symmetric'' (i.e. solutions with homogeneity
and isotropy along three spatial dimensions) five-dimensional 
spacetimes with a scalar field and a three-brane representing our universe.
We write Einstein's equations in a conformal gauge, using light-cone 
coordinates. We obtain explicit solutions: a. assuming proportionality  
between the scalar field and the logarithm of the (bulk) scale factor; b. 
assuming separable solutions. We then discuss the cosmology in the brane 
induced by these solutions. 
\end{abstract}

\section{Introduction}

The idea that our world might be a brane embedded in a higher dimensional 
spacetime has generated lately an intensive research, notably in cosmology. 
In general, the confinement of matter on the brane, leads to a cosmological 
evolution in the brane which is different from the usual evolution 
governed by Friedmann's law \cite{bdl99}. 

Such a deviation is problematic for  the ``recent'' history of the universe, 
since the usual nucleosynthesis scenario would not work anymore. The simplest
way to cure this problem \cite{cosmors}
is to introduce a negative cosmological 
constant in the bulk, \`a la Randall-Sundrum \cite{rs99b}, 
and usual cosmological 
evolution is recovered at late times when the energy density of cosmological 
matter $\rho$ 
is much smaller than the brane tension $\sigma$ (fine-tuned so as to compensate
the bulk cosmological constant) \cite{bdel99,cosmors2}.

This solution, however, requires a fine-tuning between the brane 
tension and the bulk cosmological constant and it would be desirable 
to obtain models where this requirement can be evaded.
This suggests the study of more complicated models 
containing dynamical fields in the bulk.  

As a first step, it is natural to consider the presence 
of  a scalar field in the bulk. 
This possibility has already been investigated  in several works, for various 
motivations. One of the first motivations to introduce a bulk scalar field  
was to stabilize \cite{stabilization} the distance between the two branes 
in the context of  the first  model introduced by Randall and Sundrum 
\cite{rs99a}.
A second, more recent, motivation for studying scalar fields in the bulk was
the possibility that such a set-up could provide some clue to solve 
the famous cosmological constant problem \cite{self-tuning}. 

In this perspective, some effort has been devoted to the construction 
of cosmological solutions, i.e. with time evolution in the brane, with 
a scalar field in the bulk \cite{hlz00,bicg00,bd01,fkv01} (see also the earlier 
solutions of \cite{cr99}). 
Several works have studied in particular the impact of the presence 
of a scalar field in the bulk on the cosmological evolution in the 
brane, without trying to solve the full system of equations in the 
bulk \cite{mw00,mb00,bdmp00,cm01}. However, if one wishes to evaluate    
quantitatively, and not only 
qualitatively, the impact of the bulk scalar field on the brane evolution, 
one needs in general a solution of the bulk equations. 
It is the purpose of the present work to provide some solutions of the 
full system of equations and to study the corresponding brane evolution.
The general problem being rather difficult, it will be seen that 
the requirement of obtaining exact analytical solutions is not easy 
to reconcile  with the wish to obtain realistic cosmological 
solutions. But we hope our solutions might be helpful to give some 
insights for the more general  situation.

In the present work, we have tried to obtain explicit solutions, by using 
the conformal gauge and light-like coordinates, a technique which has been 
employed for instance to the case of an empty bulk (but with  a cosmological 
constant) \cite{bcg00} or 
in the case of a bulk with a scalar field, like here, 
but with a vanishing potential \cite{hlz00}. 
We have managed to obtain explicit solutions by imposing two types of 
ans\"atze. 
A first ansatz, very powerful to integrate  the full 
system of Einstein's equations, will be  to assume
 a special relation between the 
scalar field and the bulk scale factor. As we will show, it turns out that 
the bulk solution is then necessarily static and one recovers some of the 
solutions obtained by \cite{cr99}. A second ansatz will 
be to assume that the solutions are  separable. 
In both cases, one must consider an exponential potential for the scalar
field.

The plan of the paper is the following. In the second section, we introduce 
the general set-up and write down the full system of equations. In the third
section, we study the solutions obtained with the additional constraint 
between the scalar field and the metric mentioned above. 
We then consider, in the fourth 
section, the case of separable solutions. The fifth section is devoted 
to the cosmological evolution  in the brane induced by 
the various  bulk solutions we have found. We finally conclude in the last 
section.

\section{The model}

Let us first define our set-up. We consider a scalar field $\phi$ 
living in a five-dimensional spacetime endowed with the  metric $g_{AB}$.
We also assume the existence of three-brane in this spacetime in which 
unspecified matter is confined. The action will be taken to be of the 
form 
\beq
{\cal S} = \int d^5 x \,\sqrt{-g}\,\left[ \frac{1}{2\,
\kappa^2}\,{}^{(5)}R - \frac{1}{2} \,(\nabla_A \phi)\nabla^A \phi - V(\phi)\,\right] 
+ \int d^4 x \, {\cal L}_{4},
\label{action}
\eeq
where ${}^{(5)}R$ is the scalar curvature of the five-dimensional metric 
$g_{AB}$ and 
where the second term is the action for the brane, which we leave unspecified
at this stage, but which is assumed to depend on the bulk scalar field 
in general. 

Since we consider only cosmological solutions, we will assume 
that the metric is isotropic and homogeneous along the three
ordinary spatial dimensions. Choosing  the so-called conformal gauge, 
the metric then reads 
\beq
ds^2 =g_{AB}dx^Adx^B= e^{2B(t,y)}\,(-dt^2+dy^2\,)\, + \,e^{2A(t,y)}
\, \delta_{ij}\, dx^i dx^j,
\eeq
where we have supposed flat three-dimensional subspaces for simplicity.
Introducing  the light-cone coordinates, $u \equiv t - y$ 
and $v \equiv t + y$, the metric 
can be rewritten in the form
\beq
ds^2= -e^{2B(u,v)}\,du \, dv \,+ \,e^{2A(u,v)}
\,\delta_{ij}\, dx^i dx^j \,.
\eeq

The matter content of the five-dimensional spacetime is described by 
the energy-momentum tensor, which can be derived from the action 
(\ref{action}). 
The total energy-momentum tensor is then found to  be of the form,
\beq
T_{AB}=\partial_A\phi\partial_B\phi-g_{AB}\left[{1\over 2}(\nabla_C\phi)
(\nabla^C\phi)+V(\phi)\right]+T_{AB}^{brane}, \label{mom-tensor}
\eeq
the first part coming from bulk scalar field, the second part being confined 
in the brane, which will be assumed to lie at $y=0$.

In the light-cone coordinates, the five-dimensional 
Einstein's equations, 
which  follow from the variation of the above action (\ref{action}) 
with respect to the 
five-dimensional metric, reduce, {\it in the bulk} to the following 
set of equations:
\begin{eqnarray}
A_{,uv} +3\,A_{,v}\,A_{,u}  &=& 
  \frac{{\kappa }^2}{6}\,e^{2\,B}\,V(\phi) \label{eo1}\\
- 3\,A_{,uu} +6\,A_{,u}\,B_{,u} - 3\,{A_{,u}}^2 &=&
   {\kappa }^2\,{\phi _{,u}}^2 \label{eo2}\\
- 3\,A_{,vv}+ 6\,A_{,v}\,B_{,v} - 3\,{A_{,v}}^2  &=&
   {\kappa }^2\,{\phi _{,v}}^2 \label{eo3} \\
B_{,uv} +  2\,A_{,uv}+ 3\,A_{,v}\,A_{,u}    &=&
   -\frac{1}{2}\,{\kappa }^2\,\phi _{,v}\,\phi _{,u} +
 \frac{{\kappa}^2}{4}\, e^{2\,B}\,V(\phi )\label{eo4}.
\end{eqnarray}
Here, (\ref{eo1}) is the $uv$ component of Eintein's equations,
 (\ref{eo2}) and (\ref{eo3}) are the $uu$ and $vv$ components respectively; 
(\ref{eo4}) comes from  the ordinary spatial components.
Finally, the Klein-Gordon equation in the bulk reads
\beq
\phi _{,v}\,A_{,u} + A_{,v}\,\phi_{,u} + \frac{2}{3}\,\phi_{,uv} = 
-\frac{1}{6}\,e^{2\,B}\,V'(\phi ). \label{eo5}
\eeq
Of course, there is some redundancy in the system of equations 
(\ref{eo1}-\ref{eo5}) because of Bianchi's  identities.
If one considers a free scalar field in the bulk, i.e. without 
potential, then our system of equations reduces exactly to that 
written by Horowitz, Low  and Zee \cite{hlz00}. When the potential 
is non vanishing, by contrast,  solutions such that $A=A(u)$ or 
$A=A(v)$, repectively purely outgoing and ingoing waves, are no longer 
possible, simply because  ingoing/outgoing  waves will
be scattered by the potential. 

The above system of equations applies only for the bulk and 
must now be completed by boundary conditions at the location 
 of the brane. An alternative, equivalent, 
procedure  would 
be to write down the complete Einstein and Klein-Gordon equations, with 
distributional source terms representing the brane. 
Let us start with the boundary conditions for the metric components, $A$ and 
$B$. They follow from the Darmois-Israel 
 junction conditions, which, for brane-universes, take the form 
\cite{bdl99}:
\beq
[K_{\mu\nu}]=-\kappa^2\left(S_{\mu\nu}-{1\over 3}Sg_{\mu\nu}\right),
\label{jump}
\eeq
where $K_{\mu\nu}$ is the extrinsic curvature tensor associated to the
brane, and 
where $S_{\mu\nu}$ is the four-dimensional energy-momentum tensor of the 
matter confined in the brane, which is obtained from the variation of the 
brane action with respect to the induced metric $g_{\mu\nu}$. 
Since we investigate only cosmological solutions, i.e. homogeneous 
and isotropic with respect to the three ordinary spatial dimensions, 
$S_{\mu\nu}$ is necessarily of the perfect fluid form:
\beq
S_{\mu\nu}=\left(\rho+P\right)u_\mu u_\nu+P g_{\mu\nu},
\eeq
where $\rho$ is the energy density, $P$ the pressure, and $u^\mu$ the 
unit time-like vector normal to the homogeneous and isotropic 3-surfaces.

Expressing the extrinsic curvature tensor components in terms of the 
metric components, and assuming  the usual 
mirror symmetry with respect to  the brane, the 
junction conditions (\ref{jump}) yield the two following equations,
\begin{eqnarray}
\left. \frac{\partial \, A(y,t)}{\partial\,y}\right|_{y=0}
 &=& -\frac{{\kappa }^2}{6}\,e^{B} 
\,\rho (t) \label{junc_B}\\
\left.\frac{\partial \, B(y,t)}{\partial\,y}\right|_{y=0} &=& {{\kappa }^2\over 6}\,e^B\,
   \left( 3\,P(t) + 2\,\rho (t) \right) \label{junc_A}.
\end{eqnarray}

Let us now turn to the boundary condition for the scalar field. It is related
to the variation of the brane Lagrangian with respect to the scalar field.
An explicit boundary condition for $\phi$ thus requires the 
specification of the dependence of ${\cal L}_{4}$ with respect to $\phi$. 
Several possibilities  have been 
considered in the literature (see \cite{cm01} for a detailed discussion). 
A first possibility (see e.g. \cite{mw00}) 
is to assume that the brane matter is minimally coupled 
to a metric $\tilde g_{\mu\nu}$, which is conformally related to the 
metric defined in the bulk action, i.e. 
\beq
{\tilde g_{\mu\nu}}=e^{2 k(\phi)} g_{\mu\nu}.
\eeq
In this case, one can define an energy-momentum tensor different from 
$S_{\mu\nu}$ by taking the variation of the brane action with respect to 
 $\tilde g_{\mu\nu}$ instead of $g_{\mu\nu}$, namely
\beq
{\tilde S}^{\mu\nu}={2\over \sqrt{-{\tilde g}}}{\delta 
 S_m\over \delta {\tilde g}_{\mu\nu}}.
\eeq
The corresponding energy density $\tilde 
\rho$ and $\tilde P$ are related to the former ones by the relations
\beq
\rho=e^{4k(\phi)}\tilde\rho, \qquad P=e^{4k(\phi)}\tilde P.
\eeq
Variation of the global action with respect to $\phi$ leads to a Klein-Gordon
equation of the form 
\beq
\nabla_A \nabla^A\phi={d V\over d \phi}-e^{-B} k'(\phi)S\, \delta(y),
\eeq
where $S\equiv -\rho+3P$ is the trace of the energy-momentum tensor. 
This means that the scalar field must satisfy the equation (\ref{eo5}) 
given above in the 
bulk as well as the boundary condition at the brane location $y=0$:
\beq
\left.\frac{\partial \, \phi(y,t)}{\partial\,y}\right|_{y=0}=-\frac{1}{2}\, 
e^B\, k'(\phi)\, S.
\eeq
For simplicity, we will assume in the following that 
$k(\phi)=\chi \tilde\phi$, where we have introduced the dimensionless
rescaled scalar field,
\beq
{\tilde \phi}={\kappa\over \sqrt{3}}\phi,
\eeq
which it will often be convenient to use. 

Another possibility (see e.g. \cite{mb00}), 
named volume-element coupling in  \cite{cm01}, 
is to consider an overall function of $\phi$ that multiplies the Lagrangian 
density so that 
\beq 
{\cal L}_{4}=e^{4\chi\tilde \phi} {\cal L}(g_{\mu\nu},\psi_i), 
\label{lag_matter}
\eeq
where the $\psi_i$ are all matter fields that are confined in the brane. 
In practice, in order to study the cosmological evolution in the brane, one 
wishes an effective description of the matter in the brane and a 
perfect fluid is quite appropriate for this purpose. However, a perfect 
fluid is an {\it effective} matter for which there is no natural 
fundamental description in terms of a Lagrangian density. There is 
nevertheless a substantial literature on a Lagrangian description 
of relativistic 
perfect fluids and various formulations exist. 
If all these formulations yield the same equations of motion, they are not 
equivalent at the Lagrangian level. This means, 
starting from (\ref{lag_matter}), different formulations will yield 
different couplings between the perfect fluid matter and the bulk scalar 
field. For instance, \cite{mb00} have used a Clebsh-Jordan formulation 
where the Lagrangian density is given by the pressure, whereas 
\cite{hlz00} have used a Taub type approach where the Lagrangian 
density is proportional to the energy density. 
These formulations are equivalent only when the equation 
of state is $P=-\rho$, i.e. if the  matter behaves like a cosmological 
constant, but they give different couplings  in all other cases.

In practice, we will use a boundary condition of the form
\beq
\left. e^{-B}\frac{\partial \, \tilde \phi(y,t)}{\partial\,y} \right|_{y=0} =  
 {\kappa^2\over 6}\, \gamma \,\rho (t), 
\label{junc_Phi}
\eeq
which involves all cases where the equation of state for matter is 
of the form $P=w\rho$, with $w$ constant. 
In the case of conformal coupling, the expression for $\gamma$ is given by
\beq
\gamma=-(3w-1)\,\chi. 
\label{conformal}
\eeq
In the case of volume-element coupling, one would get
\beq
\gamma=4\chi,
\eeq
if the Lagrangian density for the perfect fluid 
is proportional to the energy density as in \cite{hlz00}, 
or
\beq
\gamma= - 4 w\chi,
\eeq
if the Lagrangian density is proportional to the pressure as was 
chosen in \cite{mb00} or \cite{cm01}.
If the matter content of the brane behaves like a cosmological constant,
i.e. $w=-1$, the three above expressions 
yield, as expected,  the same result, and there is no ambiguity.

\section{``Proportional''  solutions}

In this section, we will solve the Einstein and Klein-Gordon 
equations with the assumption that 
\beq
\F(u,v)\,= \lambda\,A(u,v) \label{lambda}.
\eeq
Substituting this ansatz in the $uv$ component of Einstein's equations 
(\ref{eo1})
as well as in the Klein-Gordon equation (\ref{eo5}), one 
finds that the bulk potential for the scalar field is necessarily of the 
exponential form with  
\beq
 V(\phi)\, =\,  V_0 \, e^{-2\lambda\,\F}.
\eeq

\subsection{Bulk equations}

The $(uu)$ and $(vv)$ components of Einstein's equations, i.e. (\ref{eo2})
and (\ref{eo3}), after substitution of (\ref{lambda}), 
yield the following relations: 
\begin{eqnarray}
2B_{,u}\,&=&\,\left(1+\lambda^2\right)\, A_{,u} + \,\frac{A_{,uu}}{A_{,u}}
\label{dBu} 
\\
2B_{,v}\,&=&\,\left(1+\lambda^2\right)\, A_{,v} + \frac{A_{,vv}}{A_{,v}}.
\label{dBv}
\end{eqnarray}
These equations imply that $A(u,v)$ is necessarily of the 
form 
\beq
A(u,v)=f(\U(u)+\V(v)),
\eeq
where $f$, $\U$ and $\V$ are arbitrary functions of a single variable. 
The integration of the two above differential equations (\ref{dBu}) and
(\ref{dBv})  then gives  
\begin{eqnarray}
B(u,v)\,&=&\,\frac{1}{2}\left(1+\lambda^2\right)\, A(u,v) + \frac{1}{2}\,\ln 
\left|f'(\U(u)+\V(v))\right| \nonumber\\
&+& \frac{1}{2}\,\ln \left| \U'(u)\right| 
+ \frac{1}{2}\,\ln\left| \V'(v)\right| + \zeta \label{A},
\end{eqnarray}
where $\zeta$ is a constant of integration, 
which can be chosen arbitrarily by appropriate rescaling 
(we will take $e^{2\zeta}=4$ for convenience). 
The function $f$ is  determined by reinserting (\ref{A}) into 
the $(uv)$ component of Einstein's  equations, equation (\ref{eo1}), and 
one gets  
\beq
\U'(u)\, \V'(v)\, \left(f'' + 3\, {f'}^2- 4\, {\cal{V}}_0 \,  f'\, 
e^{\left(1- \lambda^2 \right)\, f}  \right) = 0,
\eeq
with 
\beq
 {\cal{V}}_0\,={\K52\over 6} \, V_0.
\eeq
Taking  $\U'(u)$ and $\V'(v)$  non zero (this is necessary if $V_0\neq 0$), 
this gives a second-order 
differential equation 
for $f$, which after a first integration yields 
\beq
\frac{{\cal{V}}_0}{1 - (\lambda^2/4) }\, 
e^{\left(4 - \lambda^2 \right)\, f} + \nu\, = \, f'\, e^{3\, f},
\eeq
where $\nu$ is an integration constant. 

Once  we know $f$, we can derive  the metric everywhere
\beq
ds^2\,= -\left|\U'\, \V'\, f' \right|\,e^{f\,\left(1+\lambda^2\right)}\,du \, dv \,+ \,e^{2f}\,\delta_{ij}\, dx^i dx^j \,.
\label{metric}
\eeq
This spacetime metric can in fact be written in a 
much simpler form, which turns out to be explicitly static. This 
is similar to the derivation, in \cite{bcg00}, of the cosmological 
solutions with only a (negative) cosmological constant in the bulk,
which are in fact Schwarzschild-AdS metrics. 
In order to see that, let us  introduce 
new coordinates defined by  
\begin{eqnarray}
R&=& e^{f} \\
T &=& \U - \V.
\end{eqnarray}
A straightforward substitution in the metric (\ref{metric}) yields the manifestly 
static metric
\beq
ds^2\,= -h(R)\, dT^2 + \frac{dR^2}{g(R)}+ R^2\,\delta_{ij}\, dx^i dx^j \, ,
\label{static}
\eeq
with
\beq
h(R)\, =-\frac{{\cal{V}}_0}{ 1 - (\lambda^2/4) }\,R^2 - \nu\, R^{\lambda^2-2}
\eeq
and
\beq
g(R)\, =-\frac{{\cal{V}}_0}{1 - (\lambda^2/4) }\,R^{2- 2\,\lambda^2 } 
- \nu\, R^{-2-\lambda^2}.
\eeq
These solutions correspond to the type II solutions of \cite{cr99}, which 
were obtained by looking for moving branes in static background 
metrics. Here we did not assume beforehand staticity of the background 
like in \cite{cr99}, but 
we have shown that the ansatz (\ref{lambda}) necessarily 
leads to static solutions (by 
static here, we mean a Killing symmetry in the non-trival two dimensions, 
including the case of a spacelike Killing vector if $T$ is 
a space coordinate).
This result is not so surprising since, with our ansatz (\ref{lambda}), we 
have somehow `frozen' the configuration of the scalar field  with 
respect to the (bulk) scalar field. 
Note that, for $\lambda=0$,  one  recovers  the standard result of 
brane cosmology without scalar field, $h(R)=g(R)=-{\cal{V}}_0R^2-\nu R^{-2}$, 
corresponding to a  AdS-Schwarzschild metric if ${\cal{V}}_0$ is negative.
The reason for this is that  
the scalar field is set to zero and its potential then acts as a cosmological 
constant.

\subsection{Brane motion and junction conditions}
The ansatz (\ref{lambda}) confronted to  the junction conditions 
(\ref{junc_B}) and (\ref{junc_Phi}) implies the following constraint 
on the matter coupling:
\beq
\gamma=-\lambda.
\eeq
It is now instructive to consider the junction conditions in the
static coordinate system (\ref{static}). 
The inconvenience of this  coordinate system 
 is that the brane cannot be considered at a fixed position. One 
must therefore  study the motion of the brane, in order 
to express properly the junction conditions, which in turn will give 
us the cosmological evolution inside the brane. 
We just give the results here since this calculation can be found in the 
literature (see e.g. \cite{cr99}). 
The motion of the brane can be  effectively described in the 
two-dimensional spacetime spanned by  the coordinates $(T,R)$,
where
 the brane behaves like a point with a trajectory given by $(T(\tau),R(\tau))$, where $\tau$ is the proper time of an observer comoving with the brane, i.e, 
\beq
d\tau^2\,= h(R)\, dT^2 - \frac{dR^2}{g(R)}.
\eeq
The components of the outward normal vector read
\beq
n^a\,=\,  \left(\frac{\dot R}{\sqrt{h\,g}},\sqrt{g +{\dot R}^2},0,0,0\right),
\eeq
where, in this subsection (and only here), a dot stands for a derivative 
with respect to the proper time $\tau$.
Applying the junction conditions, 
one ends up with
\beq
{\dot R^2\over R^2}={\kappa^4\over 36}\rho^2- {g(R)\over R^2}
={\kappa^4\over 36}\rho^2+\frac{{\cal{V}}_0}{1 - (\lambda^2/4) }
\,R^{- 2\,\lambda^2 } 
+ \nu\, R^{-4-\lambda^2}.
\eeq
Note that here, although the background metric turns out to be the same as 
that found in \cite{cr99}, we consider more general cosmological situations
because we do not restrict the equation of state of the matter on the brane 
to be that of a cosmological constant. In this picture, a different 
equation of state corresponds to a different trajectory in the static 
background spacetime.

\section{Separable solutions}

In this section we  look for exact solutions of Einstein's equations, 
which are separable when expressed in the light coordinate system, i.e. 
such that 
\begin{eqnarray} 
A(u, v) &=& A_1(u) + A_2(v),  \label{tina1}\\
B(u, v) &=& B_1(u) + B_2(v), \label{tina2}\\
\phi(u, v) &=& \phi_1(u) + \phi_2(v) \label{tina3}.
\end{eqnarray}
To take advantage of this ansatz, one must consider an exponential 
potential 
\beq 
V(\phi)=V_0 \, e^{\alpha\phi}=V_0 \, e^{\a\F}
, \qquad \a\equiv{\sqrt{3}\over\kappa}\alpha, \label{potential}
\eeq
 so that the potential is the product of a function of $\phi_1$ with a 
function of $\phi_2$.

\subsection{Bulk equations}

With the above ansatz (\ref{tina1}-\ref{potential}), 
all the second crossed derivatives in the bulk 
equations of motion disappear, and the system  of  differential equations 
(\ref{eo1}-\ref{eo5}) reduces to 
\begin{eqnarray} 
{A_1}'\, {A_2}'\, &=& \frac{\kappa^2 \, V_0}{18} \,e^{2\,B_1 +2\,B_2
+\alpha\,\phi_1+\alpha\,\phi_2}, \label{eo1s}\\
2\,{A_1}'\,{B_1}' - {{A_1}'}^2 - {A_1}'' - \frac{\kappa^2}{3}\,{\phi_1'}^2 
&=&0, \label{eo2s}\\
2\,{A_2}'\,{B_2}' - {{A_2}'}^2 - {A_2}'' -  \frac{\kappa^2}{3}\,{\phi_2'}^2 
&=&0, \label{eo3s}\\
6\,{A_1}'\,{A_2}' + \kappa^2\,\phi_1'\,\phi_2' & =& \frac{\kappa^2 \, V_0}{2}
e^{2\,B_1 + 2\,B_2 + \alpha \,\phi_1 + \alpha \,\phi_2}, \label{eo4s}\\
{A_2}'\,\phi_1' + {A_1}'\,\phi_2' & = & -\frac{\alpha\,V_0}{6}\, 
e^{2\,B_1 + 2\,B_2 + \alpha \,\phi_1 + \alpha \,\phi_2},\label{eo5s}
\end{eqnarray}
where the primes stand for ordinary derivatives, with respect to $u$ 
for quantities labelled by $1$ and with respect to $v$ for quantities 
labelled with $2$. It will be assumed from now on that $V_0$ is non zero. 
Comparison of (\ref{eo1s}) with (\ref{eo4s}) implies 
\beq
{A_1}'\, {A_2}'\,={\kappa^2\over 3}\phi_1'\,\phi_2',
\eeq
hence the decomposition
\begin{eqnarray} 
{\F_1}'(u) &=& C^{-1}\,{A_1}'(u), \label{F1p}\\
{\F_2}'(v) &=& C\,{A_2}'(v), \label{F2p}
\end{eqnarray}
where $C$ is a (non zero) separation constant (note that the cases 
where ${\F_1}'=0$ or ${\F_2}'=0$ are possible only if $V_0=0$). 
It turns out that this constant  cannot be chosen arbitrarily. Indeed, 
substitution of (\ref{F1p}-\ref{F2p}) into (\ref{eo5s}) and comparison 
with (\ref{eo1s}) imposes the following relation (if $V_0$ is non zero):
\begin{eqnarray} 
\tilde \alpha  = -\frac{1 + C^2}{C} \label{tinita},
\end{eqnarray}
which implies that $|\tilde \alpha|\geq 2$.
The first equation of the system, (\ref{eo1s}),  can also be separated, 
giving the two following relations,
\begin{eqnarray}
e^{2B_1}&=&De^{-\a\F_1}\F_1', \label{A1}\\ 
e^{2B_2}&=&{18\over \kappa^2 V_0 D}e^{-\a\F_2}\F_2',\label{A2}
\end{eqnarray}
where $D$ is a (non zero) 
separation constant, and where the $A_i$  have been replaced 
by the $\F_i$ using (\ref{F1p}-\ref{F2p}).
It is then easy to check that the remaining equations  
(\ref{eo2s}) and (\ref{eo3s}) are automatically satisfied, since after 
use of (\ref{F1p}) and (\ref{F2p}), they are simply the derivatives 
of the expressions (\ref{A1}) and (\ref{A2}). 

We have thus solved completely the Einstein equations in the bulk. Our 
solutions are parametrized by two arbitrary functions $A_1$ and 
$A_2$, from which all the components of the metric are obtained.
Let us now turn to the boundary conditions at the brane location.

\subsection{Junction conditions}

We must now make the link between the metric in the bulk and the 
matter in the brane. This comes from the junction conditions 
(\ref{junc_B}-\ref{junc_A}) and (\ref{junc_Phi}). 
Comparison between the junction conditions (\ref{junc_B}) and 
(\ref{junc_Phi}) leads to the following relation
\beq
A_2'(t) \,=\, \beta A_1'(t), \label{junction1}
\eeq
with 
\beq 
\beta\equiv 
\frac{C\,\gamma  + 1}
{C\,\left(C+ \gamma \right)} \label{ondinas}, 
\eeq
which can be integrated immediately if one assumes that $\gamma$ is constant, 
i.e. that $w$ is constant (note that $C+\gamma\neq 0$ otherwise 
$A_1'=0$ which is forbidden for $V_0\neq 0$).

Comparison of the junction condition (\ref{junc_A}) for $B$ 
with the junction condition (\ref{junc_B}) for $A$, in which 
one replaces $B_1$ and $B_2$ with their respective 
expression in terms of $A_1$ and $A_2$ according to 
(\ref{A1}-\ref{A2}) and (\ref{F1p}-\ref{F2p}), 
leads to a more complicated 
expression involving the second derivatives of $A_1$ and $A_2$, which 
reads
\beq
\left[-\frac{\tilde \alpha}{C}+2\left(3w+2\right)\right]A_1'
- \left[-\tilde \alpha \, C +2\left(3w+2\right)\right]A_2'=
-{A_1''\over A_1'}+{A_2''\over A_2'}, \label{junction2}
\eeq
where the $A_i$ are here only functions of time $t$ because we are on 
the brane, i.e. at $y=0$. 
This expression can be rewritten in the form 
\beq
\left[\left(-\frac{\tilde \alpha}{C}+2\left(3w+2\right)\right)
- \left(-\tilde \alpha C +2\left(3w+2\right)\right)\beta\right]A_1'=
{\beta'\over \beta}. \label{dbeta}
\eeq
For simplicity, we will restrict ourselves to the cases where the equation 
of state is such that $w$ is constant, which implies in turn that 
 $\gamma$ is constant. In this case,   the right hand side of 
(\ref{dbeta}) must vanish 
because of (\ref{ondinas}), and one 
finds  the following relation between the constants $C$, $\gamma$ and $\alpha$: 
\beq
\left( C^2 - 1\right) \,\left(\tilde \alpha \,\gamma + 2\,\left( 2 + 3\,w \right)\right)=0, \label{racataca}
\eeq
where one must recall that $\tilde \alpha$ is a function of $C$, given 
in  (\ref{tinita}). 
This condition can be satisfied 
if  $C=\pm 1$ or if  
\beq
\gamma \, =\, -\frac{2\,\left(2+3\,w\right)}{\tilde \alpha }.
\label{cond_gamma}
\eeq

As it will be  clear below  (see equation( \ref{rho})),  
 the cases $C=\pm 1$ 
(corresponding to $\a = \mp 2 $  by (\ref{tinita})) are not very interesting 
since they lead to a zero energy 
density in the brane, which means that 
 the brane is only virtual and does not really exist 
physically. For the, more interesting,  other cases, 
the condition we obtain tells us that for a given potential, i.e. for a 
given choice of the coefficient $\tilde \alpha$ (with $|\a|\geq 2$), the 
coupling between the bulk scalar field and the brane matter is enterely 
determined by the equation of state if one wishes to find separable 
solutions. 

After having established  all conditions required by  the 
compatibility between the three junction conditions, 
one can now compute the various quantities on the brane in terms of 
$A(t)$ only. 
 Indeed, starting from (see equation (\ref{junction1})) 
\beq
A(t)=(1+\beta)A_1(t),
\eeq
where the integration constant can be ignored, up to a coordinate rescaling, 
one finds for the scalar field, using (\ref{F1p}-\ref{F2p}) (and ignoring
once more the integration constants which can be reabsorbed in a rescaling 
of $V_0$), 
\beq
\F(t)=C^{-1}A_1(t)+C A_2(t)={2-\g\a \over 2\g-\a}A(t).
\eeq
The other metric component follows from (\ref{A1}-\ref{A2}) is given 
by
\beq
e^B(t)=  \sqrt{18\beta\over \kappa^2 V_0}\exp\left[-{\a\over 2}
(C^{-1}+\beta C) A_1\right] |A_1'|. \label{B}
\eeq
Finally, the energy density can be evaluated from (\ref{junc_B}). 
Inserting the product of (\ref{A1}) and 
(\ref{A2}), one finds 
\beq
{\kappa^2\over 6}\rho=e^{-B}\left(1-\beta\right)A_1'=
\pm \sqrt{\kappa^2 V_0\over 18\beta}(1-\beta) 
e^{\a\F/2}, \label{rhoA1}
\eeq
which eventually gives 
\beq
\rho \,=\pm \sqrt{ {2V_0 (\a^2 -4 )\over\K52 ( 1-\a\g\ +\g^2)}} 
\,e^{\a \F/2}\label{rho}.
\eeq
We will explore in more details the corresponding brane cosmology in the
next section.

\section{Cosmological evolution in the brane}
The purpose of this section is to analyse the cosmological behaviour 
inside the brane, given the global solutions obtained in the previous 
section.
The brane being assumed to stay at $y=0$, the induced metric in 
the brane is simply
\beq
ds^2=-e^{2B(t,0)} dt^2+e^{2A(t,0)}\d_{ij}dx^idx^j.
\eeq
The cosmological scale factor thus corresponds to
\beq
a(t)=e^{A(t,0)},
\eeq
and the cosmic time $\tau$ can be derived from the time $t$ by
\beq
\tau=\int e^{B(t,0)}dt. \label{tau}
\eeq
  
\subsection{Generalized Friedmann equations}

As a first step, we will combine some of the Einstein
equations evaluated on the brane with the junction conditions, and thus 
establish a generalized Friedmann's equation as well as a generalized 
conservation equation. In order to do so, we will closely follow  the 
derivation of \cite{bdel99} (and use the notation 
$a(t,y)\equiv e^{A(t,y)}$) with the additional ingredient that 
we allow here for 
  an energy flux from the fifth dimension, i.e. the component 
$(0,5)$ of the bulk energy-momentum tensor is non zero because of the 
presence of the scalar field. We do not rewrite here the components 
of the Einstein tensor in the $(t,y)$ coordinate system but we refer 
the reader to \cite{bdel99} where they are explicitly written.  

Substitution of the  junction conditions in the $(0,5)$ component of the 
Einstein's equations evaluated at the brane, immediately yields 
the generalized conservation equation
\beq
\dot \rho+3{\dot a \over a}(\rho+p)={\cal F},
\eeq
with 
\beq
{\cal F}=
\left. 2e^{-B} T_{05}\right|_{y=0}.
\eeq
Here $T_{05}=\dot \phi\phi'$, and using  the junction condition 
 for the scalar field, one ends up 
with the equation 
\beq
\dot \rho+3{\dot a \over a}(\rho+P)=\g\dot\F\rho.
\eeq
This generalizes the usual conservation law for cosmological matter, for wich
the right hand side is zero. The integration of the equation yields the 
following evolution for the energy density for an equation of state 
$P=w\rho$ with $w$ constant:
\beq
\rho\propto a^{-3(1+w)}e^{\g\F},
\eeq
and one recovers the familiar evolution of standard cosmology only if 
the scalar field is constant in time. 

Let us now consider the $(5,5)$ component of the Einstein equations. 
Using the $(0,5)$ component, it can be rewritten in the form 
\beq 
\dot F={2\over 3}\dot a a^3\kappa^2 T^5_5
-{2\over 3} a' a^3\kappa^2 T^5_0, \label{einstein55}
\eeq
with 
\beq
F\equiv e^{-2B}\left[{(aa')^2}-{(a\dot a)^2}\right].
\eeq
This corresponds to a slight generalization of the expression given 
in \cite{bdel99}, allowing here for a non vanishing flux from the 
extra-dimension. 
Taking the value of (\ref{einstein55}) at the brane, and using the junction
condition, one finds after integration in time, 
the following generalized Friedmann's equation
\beq
H^2={\kappa^4\over 36}\rho^2
- {2\over 3 a^4}\int d\tau  \left({da\over d\tau}\right) a^3\kappa^2 T^5_5
-{\kappa^4 \g\over 18 a^4}\int d\tau a^4 \left({d\F_0\over d\tau}\right) 
\rho^2, \label{fried}
\eeq
where the Hubble parameter is defined by 
\beq
H\equiv {da/d\tau\over a}=e^{-B}\, {\dot a \over a}.
\eeq
This equation can be found, with  a slightly different presentation, 
in \cite{bdmp00}. It is characterized by the quadratic appearance of the 
energy density of the brane, which is a generic feature of brane cosmology
\cite{bdl99}. We also have an integral term related to the {\it pressure}
along the fifth dimension and an integral term related to the energy flux 
coming from the fifth dimension, which here is essentially related to the 
time variation {\it on the brane} of the bulk scalar field. 

Finally, the expression for the energy-momentum tensor of the 
scalar field gives us
\beq
T_5^5={1\over 2}e^{-2B}\left(\phi'^2+\dot\phi^2\right)-V(\phi),
\eeq
which yields 
\beq
\left.{2\over 3}\kappa^2\,{T_5^5}\right|_{y=0}={\kappa^4\over 36} \gamma^2\rho^2+
\left({d\tilde\phi_0\over d\tau^2}\right)^2-{2\over 3}\kappa^2V_0 e^{\alpha
\tilde\phi_0},
\eeq
 where we have used the junction condition (\ref{junc_Phi}) to replace 
the gradient contribution by the term quadratic in the energy density.

\subsection{Static bulk solutions}

The generalized Friedmann equation
has the form 
\beq
H^2={\kappa^2\over 36}\rho^2 +\,\frac{{\cal{V}}_0}{1 - (\lambda^2/4) }
\,a^{-2\lambda^2}+\,\nu\,a^{-4-\lambda^2}.
\label{friedstatic}
\eeq
Combining this Friedmann equation with the conservation equation, 
one can obtain explicitly the evolution of the scale factor as a function 
of the cosmic time for any given equation of state. 

It is easy to check that the general form (\ref{fried}) for the 
Friedmann equation is compatible with the above Friedmann equation. 
The five-dimensional pressure can be expressed in the form
\beq
\left.{2\over 3}\kappa^2\,{T_5^5}\right|_{y=0}={\kappa^4\over 18} \lambda^2\rho^2-
\lambda^2 {g(a)\over a^2}-{2\over 3}\kappa^2V_0 a^{-2\lambda^2},
\eeq
where we have used the Friedmann's equation in the form (\ref{friedstatic}) 
to replace the 
kinetic term. 
Substituting in the integral terms, one finds that the two $\rho^2$ terms
just cancel, and after integration of the remaining terms, one obtains 
directly what one expects, i.e. 
\beq
{2\over 3 a^4}\int d\tau  \dot a a^3\kappa^2 T^5_5
+{\kappa^4\over 18 a^4}\int d\tau a^4 \g\dot\F \rho^2
= {g(a)\over a^2}.
\eeq

\subsection{Separable solutions}

According to (\ref{tau}) and (\ref{B}), the cosmic time is given by
\beq
\tau=\sqrt{\frac{18\,\beta}{\kappa^2\,V_0}}\,\mu^{-1}\,e^{\mu A_1}
\equiv \tau_0 \,e^{\mu A_1},
\qquad
\mu={\a\,(\a\g-2)\over 2\,(\g+C)},
\eeq
where we have assumed $A_1'>0$ for definiteness. 
The scale factor is given by
\beq
a(\tau)=\left({\tau\over \tau_0}\right)^p, \qquad p\equiv {1+\beta\over 
\mu}= {2 \,(\a-2\g) \over \a\,(2-\g\a)}.
\eeq
Substituting the condition (\ref{cond_gamma}) for the coupling $\gamma$, 
in the interesting case where the brane is not empty, the 
power can be written
\beq
p={1\over 3(1+w)}\left[1+{4(2+3w)\over\a^2}\right].
\eeq
The minimum value of $\a^2$ being $4$, the range of possible $p$ 
is between $p=1$ and $p=1/(3(1+w))$, the latter being the unconventional 
evolution in the case of an empty bulk \cite{bdl99}. Note that this range 
includes the conventional value $p=2/(3(1+w))$.
    
The Hubble parameter is therefore
\beq
H={p\over \tau}.
\eeq
The energy density in the brane can be related to this Hubble paramter, 
the easiest way being to use (\ref{rhoA1}) and one finds that the 
two are proportional,
\beq
{\kappa^2\over 6}\rho={1-\beta\over 1+\beta}H. 
\eeq
The unconventional cosmological evolution that was obtained in the case
of an empty bulk \cite{bdl99} is thus generalized here for a bulk 
with a scalar field for our specific separable solutions. There is 
a difference however is the proportionality coefficient.
Finally, the evolution of the scalar field is given 
by 
\beq
\F_0={2-\g\a \over 2\g-\a}\ln a. \label{cachislamar}
\eeq

It is now instructive to 
check  the generalized Friedmann equation (\ref{fried}), 
by  evaluating   the 
contribution of each term for our explicit solutions.  
One  finds 
\beq
-{2\over 3 a^4}\int d\tau  \left({da\over d\tau}\right)
 a^3\kappa^2 T^5_5 = -2 {\a^2\g^2 + 4\a\g -8\g^2 -4 \over (\a -2\g)(\a^2\g +2\a -8\g)} \frac{p^2}{\tau^2}
\eeq
and 
\beq
-{\kappa^4 \g\over 18 a^4}\int d\tau a^4 \left({d\F_0\over d\tau}\right)
   \rho^2 = - 2\g { (\a^2 -4) (\a\g -2) \over (\a -2\g)^2 (\a^2\g +2\a -8\g)} \frac{p^2}{\tau^2}
\eeq
and the sum of these two terms indeed coincides with 
\beq
H^2-{\kappa^4\over 36}\rho^2 = 4 {1-\a\g+\g^2 \over (\a-2\g)^2} H^2.
\eeq
These expressions can be rewritten, using (\ref{cond_gamma}), in terms
of the parameter $w$ of the equation of state. One finds
\beq
-{2\over 3 a^4}\int d\tau   \left({da\over d\tau}\right) a^3\kappa^2 T^5_5 
= 4{(9w^2+6w-1)\a^2 -8(9w^2+12w+4)\over (8+12w+\a^2)
((3w+1)\a^2 -8(3w+2))}\frac{p^2}{\tau^2},
\eeq 
\beq
-{\kappa^4 \g\over 18 a^4}\int d\tau a^4 \left({d\F_0\over d\tau}\right)
  \rho^2 = 12\a^2{(1+w)(2+3w)(\a^2-4) \over  (8+12w+\a^2)^2 ((3w+1)\a^2 
-8(3w+2))} \frac{p^2}{\tau^2},
\eeq
and 
\beq
H^2-{\kappa^4\over 36}\rho^2 = 4 
{(5+6w)\a^2+4(2+3w)^2 \over  (8+12w+\a^2)^2} H^2.
\eeq

Let us also mention the particular case  
$\gamma \a= 2$, corresponding to the equation of state $w =-1$. 
One can easily see that the energy density is constant, given by   
\beq
\rho =\sqrt{ -{4 \a^2 V_0\over \K52}}, 
\eeq 
which implies in particular that the scalar field potential must be 
negative for consistency. The scale factor, is given, instead of a power-law,
by an exponential, namely
\beq
a(\tau) = \exp\left[\sqrt{-{\K52 V_0 \over 18}(\a^2-4)} \,\, \tau\right].
\eeq

\subsection{Cosmology with conformal coupling}

In the case of conformal coupling, one must be aware that the cosmological 
evolution given above corresponds to the so-called Einstein frame, i.e. 
the frame associated with the metric for which the action is the usual 
Einstein-Hilbert action (here we extend this definition to the five 
-dimensional gravitational action). However, the physical metric corresponds
to the metric which is minimally coupled to the ordinary matter and in our 
case, this means that the physical frame, usually called the Jordan 
frame in the context of scalar-tensor theories of gravitation, is the frame 
associated with the metric $\tilde g_{\mu\nu}$.

It is therefore useful to rewrite the previous equations in the physical 
frame. The correspondance between the two scale factors is 
\beq
{\tilde a}=e^{\chi\F}a.
\eeq
The proper times are also different and related by 
\beq
d{\tilde \tau}=e^{\chi\F} dt.
\eeq
Using the relation (\ref{cachislamar}), one finds that 
\beq
e^\F=\left({\tau\over \tau_0}\right)^{-2/\a},
\eeq
and therefore, the expression of the new proper time $\tilde\tau$ in terms 
of the ``Einstein frame'' proper time is 
\beq
\tilde \tau=\tilde\tau_0\left({\tau\over \tau_0}\right)^{1-{2\chi\over \a}}
, \qquad \tilde\tau_0\equiv {\tau_0\over 1-{2\chi\over \a}}.
\eeq
The cosmological evolution of the scale factor is thus given by 
\beq
\tilde a=\left({\tilde \tau\over\tilde \tau_0}
\right)^{\left(p-{2\chi\over \a}\right)/ 
\left(1-{2\chi\over \a}\right)}.
\eeq
Note that one cannot find a separable solution with conformal coupling 
in the case of radiation. Conformal coupling, according to  (\ref{conformal}), 
implies $\gamma=0$ for radiation, which is clearly incompatible with the 
condition (\ref{cond_gamma}) for separable solutions.

\section{Conclusions}
In the present work, we have obtained explicit exact solutions for 
a five-dimensional spacetime with a scalar field and a 3-brane. 
Although several explicit solutions exist already in  the literature, 
most of them correspond to static solutions for which the 
bulk geometry and scalar field are frozen and the cosmological evolution 
is only due to the motion of our brane-universe in this bulk.
As we have shown here, this is the case when one assumes some 
proportionality relation between the scalar field and the logarithm 
of the bulk scale factor. More
interestingly, we have also found  solutions where the bulk itself 
has some dynamics. Even if our  solutions are rather artificial, 
 and cannot be considered as realistic models for the 
recent cosmology of our universe, they might be useful to give some insights
in the future investigations of bulk-brane dynamics.

Although these solutions 
 might represent acceptable models for the very early universe, 
one of the main open problems 
 is  to obtain generic solutions of the 
bulk-brane system.   One could then investigate whether 
 there exist, or not,  some attractor solutions so that the primordial brane 
cosmology would be somewhat insensitive to the initial conditions. 

Another open question is how, in a cosmological context, the presence 
of the scalar field and of its perturbations might affect the 
results obtained so far for cosmological perturbations in a brane-universe,
where the bulk was assumed to be empty apart the presence of a negative 
cosmological constant (see \cite{l00} and references therein). 

\vskip 1cm

{\bf Acknowledgments}: we would like to thank S. Carroll for interesting 
discussions, as well as  C. Charmousis for his useful comments 
on the present work and for informing us about 
his work in progress, where he has obtained, independently, and using 
a similar technique, some of the results presented here.

\end{document}